\documentclass[a4paper,fleqn,usenatbib]{mnras}
\usepackage{newtxtext}

\usepackage[T1]{fontenc}
\usepackage{ae,aecompl}

\usepackage{graphicx}	
\usepackage{amssymb}	
\usepackage[hyphenbreaks]{breakurl}

\title[Characterization of high proper motion sources]{Spectro-photometric characterization of high proper motion sources from WISE}

\author[J. C. Beam\'in]{
J. C. Beam\'in,$^{1,2,3}$\thanks{E-mail: jcbeamin@astro.puc.cl}
V. D. Ivanov,$^{1,4}$
D. Minniti,$^{5,3,6}$
R. L. Smart.$^{7}$
K. Mu{\v z}i{\'c},$^{1}$
\newauthor
R. A. Mendez,$^{8,1,3}$
Y. Beletsky,$^{9}$
A. Bayo,$^{10,3}$
M. Gromadzki,$^{3,10}$
R. Kurtev,$^{10,3}$
\\
$^{1}$European Southern Observatory, Ave. Alonso de Cordoba 3107, Casilla 19001, Santiago, Chile.\\
$^{2}$Instituto de Astrof\'isica, Facultad de F\'isica, Pontificia Universidad Cat\'olica de Chile, Casilla 306, Santiago 22, Chile.\\
$^{3}$Millennium Institute of Astrophysics, Chile.\\
$^{4}$European Southern Observatory, Karl-Schwarzschild-Str. 2, D-85748 Garching bei M\"unchen, Germany\\
$^{5}$Facultad de Ciencias Exactas, Universidad Andres Bello, Fernandez Concha 700, Las Condes, Santiago, Chile\\
$^{6}$Vatican Observatory, V00120 Vatican City State, Italy\\
$^{7}$INAF-Osservatorio Astrofisico di Torino, Strada Osservatorio 20, 10025 Pino Torinese, TO, Italy.\\
$^{8}$Universidad de Chile, Departamento de Astronom\'ia, Casilla 36-D, Santiago, Chile.\\
$^{9}$Las Campanas Observatory, Carnegie Institution of Washington, Colina el Pino, 601 Casilla, La Serena, Chile\\
$^{10}$Instituto de F\'isica y Astronom\'ia, Facultad de Ciencias, Universidad de Valpara\'iso, Ave. Gran Breta\~na 1111, Playa Ancha, Valpara\'iso, Chile.\\
}

\pubyear{2015}

\begin{document}
\label{firstpage}
\pagerange{\pageref{firstpage}--\pageref{lastpage}}
\maketitle

\begin{abstract}
  The census of the solar neighborhood is almost complete for stars and
  becoming more complete in the brown dwarf regime. Spectroscopic, photometric
  and kinematic characterization of nearby objects helps us to understand the
  local mass function, the binary fraction, and provides new targets for
  sensitive planet searches.  We aim to derive spectral types and
  spectro-photometric distances of a sample of new high proper motion
  sources found with the WISE satellite, and obtain parallaxes for
  those objects that fall within the area observed by the Vista Variables in
  the V\'ia L\'actea survey (VVV).  We used low resolution spectroscopy and
  template fitting to derive spectral types, multiwavelength photometry to
  characterize the companion candidates and obtain photometric distances.
  Multi-epoch imaging from the VVV survey was used to measure the parallaxes
  and proper motions for three sources.  We confirm a new T2 brown dwarf
  within $\sim$15\,pc. We derived optical spectral types for twenty four
  sources, mostly M dwarfs within 50\,pc. We addressed the wide binary nature
  of sixteen objects found by the WISE mission and previously known high proper
  motion sources. Six of these are probably members of wide binaries, two of
  those are new, and present evidence against the physical binary nature of
  two candidate binary stars found in the literature,  and eight that we selected
  as possible binary systems.  We discuss a likely microlensing event produced
  by a nearby low mass star and a galaxy, that is to occur in the following
  five years.
\end{abstract}


\section{Introduction}

High proper motion sources have been studied for over two centuries now, as
one might expect that the fastest moving sources would be the closest to our
solar system.  Since mid XX century, comprehensive searches of high proper
motion objects have been performed using  photographic plates and later CCD cameras,
to find the closest neighbors of the Sun \citep[][RECONS, among
  others]{Giclas1971, Luyten1979a, Luyten1979b, Wroblewski1989, Hambly2004,
  Lepine2005a, Lepine2008, Finch2014}.   However, cool objects are
intrinsically faint in the optical and emit most of their light in the near
infrared (NIR). Therefore, the searches for unknown nearby sources gradually
turn to the NIR wavelengths taking advantage of improved camera sensitivities,
spatial and temporal resolution \citep{Deacon2009, Kirkpatrick2010,
  Smith2014}, yielding in the last twenty years the discovery of over two
thousand ultra-cool dwarfs. Widening the color space of the searches has
helped to improve the stellar and sub-stellar density estimates in the solar
neighborhood, and the frequency of low-mass companions, among other questions
\citep{Allen2012, Luhman2012, Dieterich2012, Ivanov2013, Deacon2014,
  Davison2015}.
 
The Wide-field Infrared Survey Explorer ({\it WISE;} \citealt{Wright2010})
satellite has revolutionized the field of brown dwarf with color based selections
providing the discovery of hundreds of brown dwarfs
\citep{Kirkpatrick2011}. Later, the multiepoch nature of the WISE mission
provided proper motions for over twenty thousand individual sources
\citep{Luhman2014a,Kirkpatrick2014}, including the discoveries of the third
and fourth closest systems to the Sun \citep{Luhman2013,Luhman2014b}.  These
are the closest binary BD, and the coldest BD known. The WISE mission also
lead to the discovery of the first Y dwarfs
\citep{Cushing2011,Kirkpatrick2012}. Thousands of other interesting objects
were found, including young very low mass objects \citep[M, L and T
  type;][]{Gagne2015}, several ultra cool sub-dwarfs \citep{Kirkpatrick2014},
etc.  These discoveries are helping to better understand the role of
temperature, metallicity and evolution in very cool atmospheres. Finally, a
new sample of nearby K and M dwarfs (d$\leq$100\,pc) was created, well suited
for exoplanet searches.

Here we report a follow up study of bright high proper motion objects detected
by \citet{Luhman2014a} and \citet{Kirkpatrick2014}, concentrating on objects
within 50\,pc, wide co-moving binaries, and possible members of nearby young
moving groups.  The project summarized here was motivated by the recent
discoveries of nearby stellar and sub-stellar objects \citep[e.g.,][ among
  others]{Artigau2010, Lucas2010, Luhman2013, Mamajek2013, Scholz2014}, the
implications that these findings may have on the stellar census in the Solar
neighborhood \citep{Henry2006,Faherty2009,Winters2015}, new results in the
multiplicity of young low-mass stars and brown dwarfs in the field and young
moving groups \citep{Delorme2012,Elliott2014}, and even the dynamic
interactions of the Solar system \citep{Ivanov2015,Mamajek2015}.  The paper is
organized as follows: Section \ref{Sec:data} describes the catalogs and
methods we used to generate a list of objects of interest, and the instruments
we used to follow up and characterize them.  In section \ref{Sec:spectra} the
methods for classification are presented. Section \ref{Sec:dist} describes the
distance measurements and comparison of photometric and spectroscopic results.
Finally, in section \ref{sec:discussion} we discuss individual sources, and
present our conclusions in section \ref{conclusions}.

\section{Sample selection and observations}
\label{Sec:data}
We started by analyzing bright sources in the new catalogs of high proper
motion sources by \cite{Luhman2014a} and \cite{Kirkpatrick2014}. We selected
the brightest sources with the highest proper motion that were visible in
April from the southern skies (i.e. Dec$<$ 30), with no previous derived
spectral types and no data in the ESO archive. We also performed a cross check
with previously known high proper motion sources from Simbad
\citep{Wenger2000}.  If Simbad returned a source within 15\arcmin\ showing a
similar proper motion, we selected the brighter one between the SIMBAD match
and the WISE object for spectroscopic follow up. We used a relaxed criterion
to select candidate companions, i.e. a source with proper motion $\geq$
200\,mas\,yr$^{-1}$ and a position angle of proper motion within
$\sim$30$^\circ$. These relaxed constraints caused the selection of some
spurious pairs, and we discuss this issue in the coming sections. We also
created a reduced proper motion diagram in order to find brown dwarf
candidates.

To derive the spectral types of the selected objects, we obtained spectra in
the optical and  in the near-infrared for WISE J21210032-6239194 (hereafter WISE 2121-6239). 
In addition, for the objects located within the
area covered by the VISTA Variables in the V\'ia L\'actea survey (VVV) we
obtained photometry and astrometry.

\begin{table}
\caption{Sample of WISE high proper motion objects selected for spectroscopic
  follow up with the EFOSC2 at the NTT at La Silla observatory, and FIRE at
  Baade at Las Campanas observatory.}            
\label{Tab:select}     
\centering                         
\begin{tabular}{@{                              }      c@{               } c@{             } c@{         } c@{          }}
\hline\hline                 
 Name      &  R.A. & DEC. & Exposures \\    
\hline             
2MASS J06571510-1446173  & 06:57:14.44 &     -14:46:25.2 &   2x120    \\ 
2MASS J07523088-4709470  & 07:52:31.93 &     -47:09:49.8 &   3x60     \\
2MASS J08291581-5850305  & 08:29:15.28 &     -58:50:35.2 &   2x60     \\   
2MASS J09432908-0237184  & 09:43:28.09 &     -02:37:22.4 &   3x60     \\
2MASS J10570299-5103351  & 10:57:01.20 &     -51:03:37.2 &   2x120    \\
2MASS J11161471-4403252  & 11:16:14.91 &     -44:03:28.4 &   2x120    \\
2MASS J11163668-4407495  & 11:16:36.94 &     -44:07:47.2 &   3x120    \\ 
2MASS J13211484-3629180  & 13:21:13.94 &     -36:29:11.1 &   6x120    \\
2MASS J13552455-1843080  & 13:55:23.50 &     -18:43:18.1 &   5x180    \\ 
2MASS J14033647+0412395  & 14:03:36.06 &     +04:12:28.9 &   5x180    \\
2MASS J14040025-5923551  & 14:04:00.52 &     -59:23:53.3 &   2x120    \\ 
2MASS J12412819-6507578  & 12:41:28.10 &     -65:07:55.6 &   2x120    \\
2MASS J13322604-6621419  & 13:32:26.19 &     -66:21:33.8 &   3x120    \\
2MASS J14035016-5923426  & 14:03:50.92 &     -59:23:45.2 &   2x120    \\
2MASS J14233830+0138520  & 14:23:38.27 &     +01:38:41.7 &   3x120    \\ 
2MASS J14574906-3904511  & 14:57:49.68 &     -39:04:55.1 &   5x180    \\
2MASS J15463089-5258367  & 15:46:31.10 &     -52:58:32.1 &   3x60     \\ 
2MASS J15480441-5810533  & 15:48:03.58 &     -58:10:49.9 &   2x120    \\ 
2MASS J17345391-6206546  & 17:34:53.20 &     -62:06:52.1 &   3x15     \\
2MASS J19104599-4133407  & 19:10:45.27 &     -41:33:46.9 &   3x120    \\ 
2MASS J19242108-0804516  & 19:24:21.09 &     -08:04:51.6 &   2x120    \\
2MASS J20044356-7123334  & 20:04:43.09 &     -71:23:28.4 &   2x120    \\ 
2MASS J21252081-3422144  & 21:25:20.78 &     -34:22:20.3 &   2x120    \\
2MASS J22275385-2337300  & 22:27:52.75 &     -23:37:35.1 &   3x120    \\ 
WISE J212100.87-623921.6 & 21:21:00.88 &     -62:39:21.7 &   6x63.4   \\
\hline                                   
\end{tabular}
\end{table}

\subsection{NTT/EFOSC2}
The ESO Faint Object Spectrograph and Camera v.2
\citep[EFOSC2;][]{Buzzoni1984,Snodgrass2008} mounted at the 3.6m New
Technology Telescope (NTT) at la Silla observatory, is a versatile instrument for low resolution
spectroscopy, imaging, and polarimetry.  We obtained low resolution long slit
spectra for twenty four sources (see Table \ref{Tab:select}) using the same
configuration for all the sources: 1\arcsec\ slit width, grism number one,
covering a spectral range of $\lambda$= 3185-10940 \AA \footnote{According to
  instrument manual, there is second order contamination for wavelengths
  longer than 9280 \AA }\, with a resolution $\sim$48 \AA. All the sources
were observed on April 17$^{th}$ 2014. The usual reduction steps, bias and
flat fields corrections were performed, followed by wavelength calibration and
flux calibration with the F type spectro-photometric standard LTT 9239
\citep{Hamuy1994}.

\subsection{Magellan/FIRE}
The Folded-port Infrared Echellette
\citep[FIRE;][]{Simcoe2008,Simcoe2010,Simcoe2013} mounted at the 6.5m Magellan
Baade Telescope uses a 2048$\times$2048 HAWAII-2RG array. It covers the
wavelength range from 0.8 to 2.5 $\mu$m when used in the high-throughput prism
mode, delivering a resolution varying from $\sim$ 500 at $J$-band to $\sim$300
at $K$-band for the 0\arcsec.6 slit.  A spectrum of the objects WISE 2121-6239
was obtained in the ABBAAB pattern, with the exposure time of $6
\times63.4$\,s. The 1\arcsec\, slit was oriented along the parallactic angle.
The read-out mode was ``Sample-Up-the-Ramp'' (SUTR) with the low gain mode
(3.8e$^-$ per count).  The A0V star HD 195288 was observed with the same set
up (with the exception of the read out mode, we used mode ``Fowler 1''
instead) in ABBA pattern, for telluric correction.  Low and high voltage (1.5
and 2.5\,V) flats were obtained after the observations of the source and the
telluric lines to correct the red and blue parts of the spectrum.  NeAr lamp spectra
were obtained after the observations for wavelength calibration. In addition,
we observed an extra lamp with the 0.45\arcsec slit to differentiate some
blends with lines that help us to improve the wavelength calibration.

The data was reduced with the instrument pipeline {\it
  firehose\_ld}\footnote{\url{http://web.mit.edu/~rsimcoe/www/FIRE/ob_data.htm}}.
It traces the slit, performs a wavelength solution, combines the flat fields
and applies them, finding the object on the 2-D image, and finally it extracts a
1-D spectrum.

The individual spectra of the target and the telluric are median-scaled with
the {\it fire\_xcombspec} tool and then combined with a Robust Weighted Mean
algorithm. Finally, the telluric correction was applied using
the {\it fire\_xtellcor\_ld}, a clone of the xtellcor program from {\it
  SpexTool} \citep{Cushing2004,Vacca2003}

\subsection{VISTA/VIRCAM}
The VVV survey \citep{Minniti2010, Saito2012, Hempel2014} is one of the six
ESO public surveys carried out with the 4.1m Visual and Infrared Survey
Telescope for Astronomy (VISTA) telescope and VIRCAM camera at cerro Paranal
Chile \citep{Dalton2006,Emerson2010}. It has sixteen 2048x2048 pixels chips
with a pixel scale of 0.34\arcsec. The total field of view is $\sim$
1$^{\circ}$ x 1.5$^{\circ}$.  The VVV is a variability survey primarily
designed to trace the 3-D structure of the Milky Way, observing 562$^{\circ2}$
towards the Galactic Bulge and southern inner disk
\citep{Gonzalez2011,Dekany2013,Minniti2014}.  This survey matches the
requirements for good estimations of the proper motions (PMs) and parallaxes,
as demonstrated previously by \citet{Beamin2013,Ivanov2013}.  The data were
reduced at the Cambridge Astronomy Survey Unit (CASU) with pipeline v1.3.  In
this paper we used the astrometric and photometric catalogs generated from
individual exposures (named pawprint, see \citealt{Hempel2014} for further
details), as opposed to catalogs generated from the combined exposures. This
way, we have two independent astrometric/photometric points obtained within a
few minutes, where the target is observed by a different detector, giving us a
better handling of the systematic uncertainties.  We took advantage of the
high stellar density of the fields and constructed a local astrometric solution to fit
relative proper motion and parallax simultaneously.

\section{Spectro-photometric  characterization}
\label{Sec:spectra}

\subsection{Spectral typing}
Twenty one of the twenty four spectra observed with EFOSC2 showed the
characteristic molecular absorption bands M type stars (TiO, CaH), and they
were compared to the primary K5V-M9V standards from \cite{Kirkpatrick1991}
and \cite{Kirkpatrick1999} available from the Dwarf
Archives\footnote{\url{http://dwarfarchives.org}}.  The remaining three
spectra indicated hotter stars, and for those earlier than K5 type objects we
used the standards of \cite{Pickles1998}.

The low resolution of our spectra ($\sim$40 \AA) does not allow to use the
spectral indexes frequently used in the literature
\citep[e.g.][]{Kirkpatrick1991,Gizis1997,Lepine2013} so we 
relied instead on the overall shape of the spectrum.

The template spectra were smoothed to the resolution of our data, normalized
at 7500 \AA.  To find the best match, we performed a $\chi^2$ minimization
over two different wavelength ranges. First for the twenty later type stars we
used the range $\lambda$=6500-9000 \AA, as most of the flux and spectral
features of interest are in that region, also the templates from
\cite{Kirkpatrick1991} that we used do not cover bluer wavelengths. For the
four hotter objects we minimize over a broader wavelength region,
$\lambda$=3500--9500\,\AA\, as they have more flux towards the bluer regions.

\begin{figure*}
\centering
   \includegraphics[width=\linewidth]{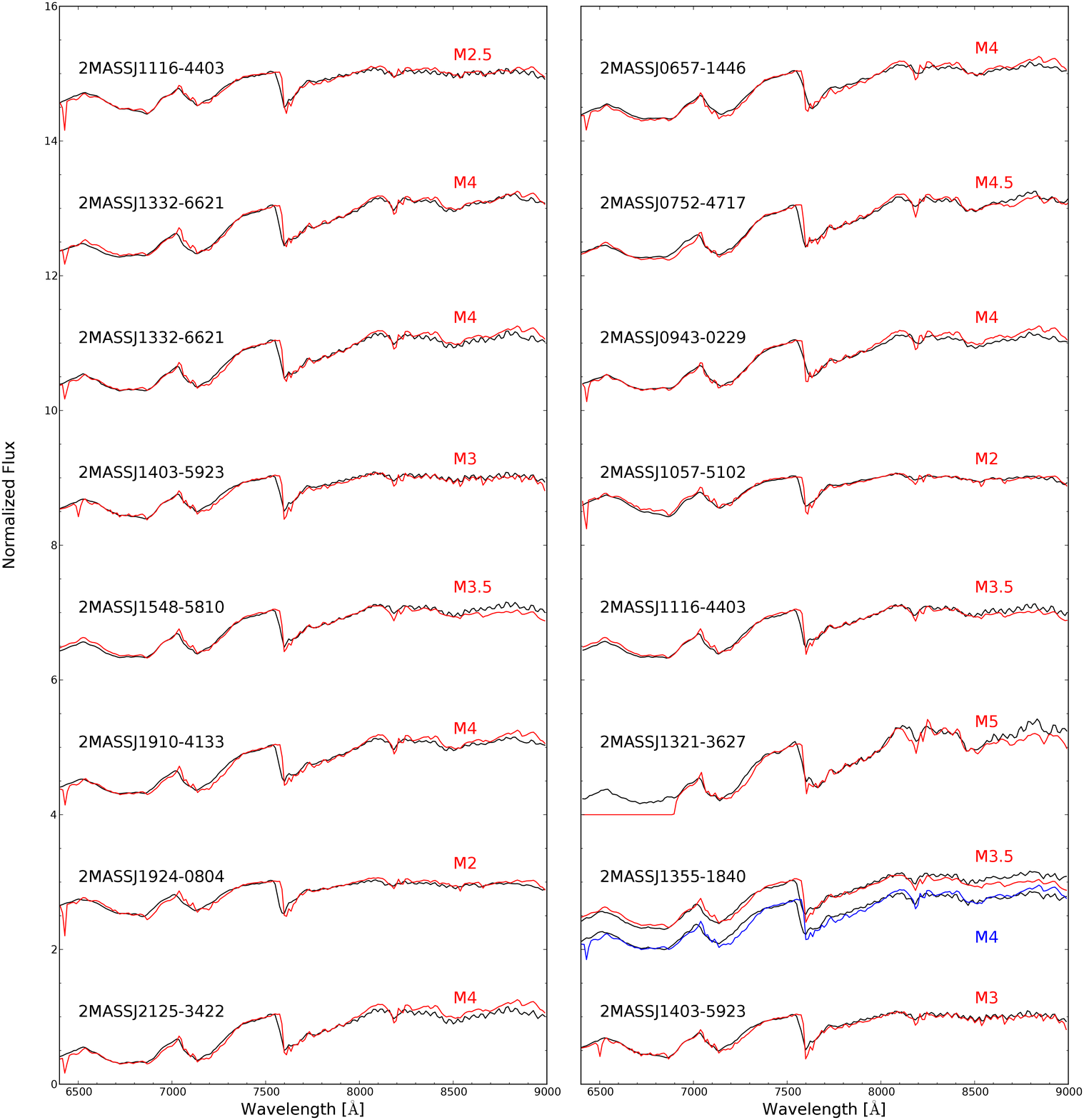}
      \caption{Each panel show the best $\chi^2$ fit to the spectrum from the
        set of spectral templates given in Table \ref{Tab:dists}, the
        templates spectra are taken from the primary standards from
        \citet{Kirkpatrick1991}. In the case of 2MASS\,1355-1840 we
        over-plotted the two best fits, as they fit slightly better different
        wavelength regions.}
         \label{sptypes1}
 \end{figure*}

\begin{figure*}
\centering
   \includegraphics[width=\linewidth]{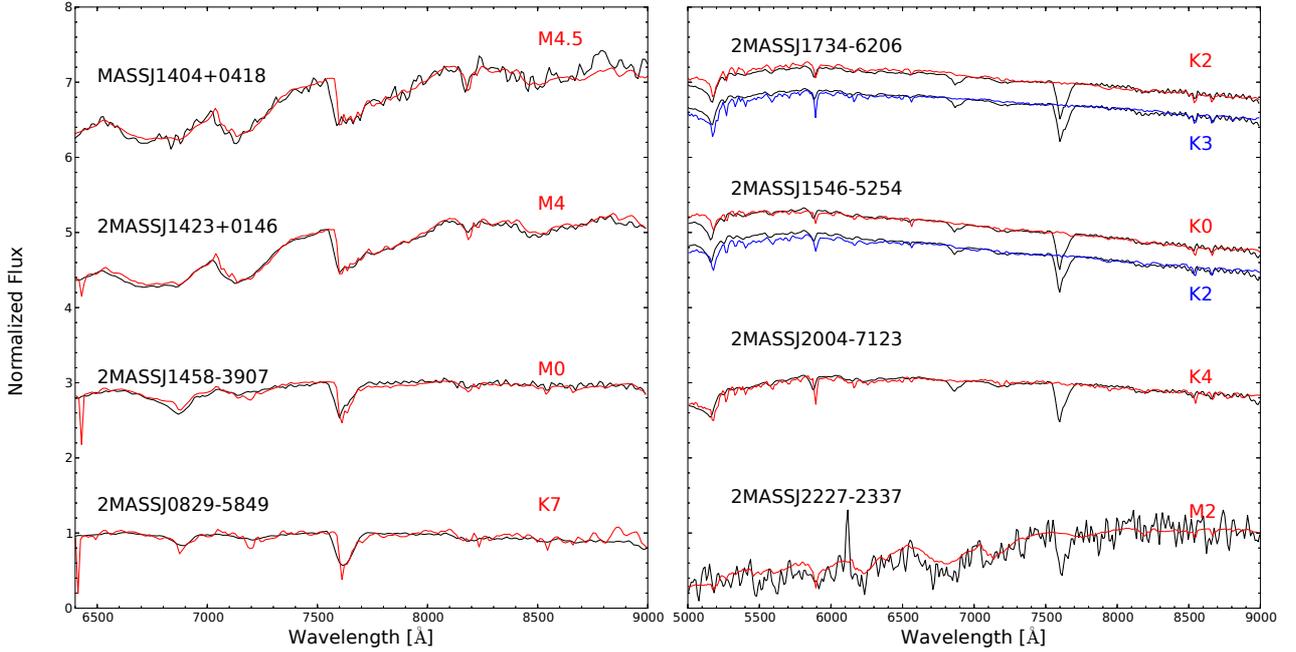}
      \caption{Left panel same as Fig \ref{sptypes1}. Right panel show the
        earlier type objects, as the \citet{Kirkpatrick1991} only goes until
        K5 we used templates from \citet{Pickles1998}.  For object
        2MASS\,1546-5252 we assume a K1 spectral type, a compromise between
        the fit to the continuum of a K0 standard and the Mg absorption around
        5170\AA \, of the K2 standard.}
         \label{sptypes2}
   \end{figure*}

The results are shown in Fig. \ref{sptypes1} and \ref{sptypes2}.  The types of
some objects were adjusted by up to 0.5 sub-type after a visual inspection.

The derived spectral types are listed in Table 3. We also attempted to fit
separately the blue (6500-7600 \AA) and red (7600-9000 \AA) parts of each
spectrum as shown in \cite{Jao2008} the red part of M sub-dwarfs
appears $\sim$1 sub-type earlier than their blue part (depending on
metallicity). We did not observe this behavior in the previously reported
sub-dwarf object 2MASS J14574906-3904511.

\begin{table*}
\caption{Main spectral indexes for T dwarfs as given in \citet{Burgasser2006}}             
\label{table:T2_index}  
\centering                         
\begin{tabular}{c c c c c c c}   
\hline\hline 
Object & H$_2$O-J & CH$_4$-J & H$_2$O-H & CH$_4$-H & CH$_4$-K & adopted type\\    
\hline 
    WISE J2121-6239 & 0.55(T2) & 0.72(T1) & 0.48(T2) & 0.93(T1) & 0.62(T2) & T2$\pm$1 \\      
\hline                
\end{tabular}
\end{table*}

\begin{figure}
\centering
   \includegraphics[width=\columnwidth]{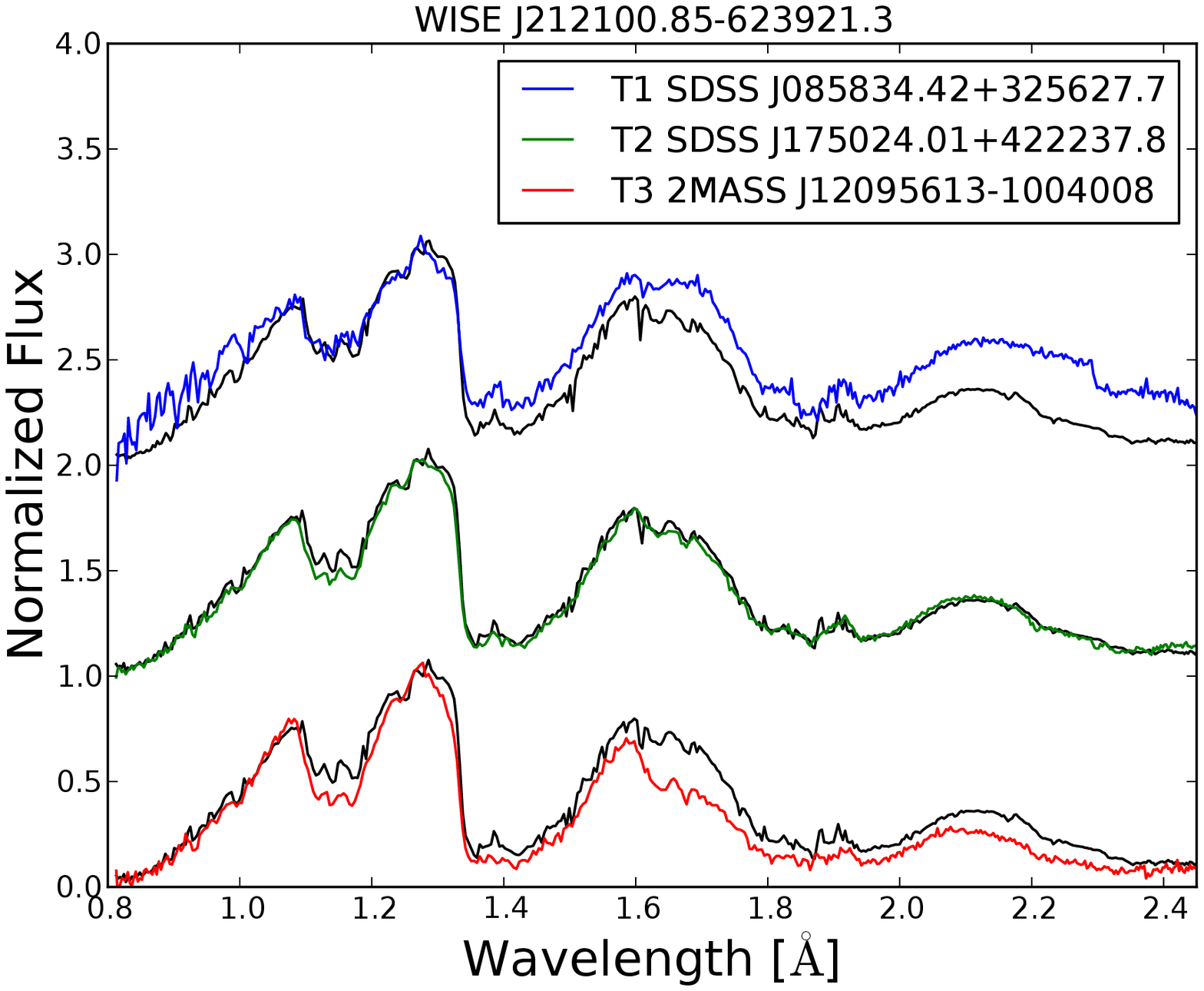}
    \caption{Black spectrum shows the new T2 dwarf WISE 2121-6239 compared to
      T1 template \citet{Burgasser2010} T2 and T3 template
      \citet{Burgasser2004}.}
\label{new_T2}
\end{figure}

The NIR spectra of WISE J2121-6239 was compared to the L and T dwarf spectra
from the Spex library 
\footnote{\url{http://pono.ucsd.edu/~adam/browndwarfs/spexprism/library.html}}
using the $\chi^2$ minimization algorithm, mentioned before, yielding a best
match to the T2 type SDSS J175024.01+422237.8. Figure \ref{new_T2} shows the
comparison of WISE 2121-6239 against three T sub-types, T1 SDSS
J085834.42+325627.7; T2 SDSS J175024.01+422237.8; T3 2MASS J12095613-1004008.
 
The spectral indexes: H$_2$O$-J$, CH$_4-J$,H$_2$O-$H$,CH$_4-H$ and CH$_4-K$
from \cite{Burgasser2006} yield a spectral type of T2$\pm$1 (Table
\ref{table:T2_index}).  Using the tabulated coefficients of Table 14 in
\cite{Dupuy2012} and the 2MASS and WISE magnitudes we computed a photometric
distance between 12-16\,pc for this brown dwarf.
   
\subsection{SED fitting}
We searched various archives for historic multi-wavelength observations of our
program targets. Using Aladin \citep{Bonnarel2000}, we were able to retrieve
the catalogs and the archival images, for visual inspection to ensure that our
high PM target cross identifications were correct.  We also used Topcat
\citep{Taylor2005} to cross-match catalogs.

To build the SED and compare to stellar models we used the Virtual Observatory
SED analyzer \citep[VOSA;][]{Bayo2008}, fitting the BT-settl models 2012
\citep{Allard2012} to the data, first with a Bayesian approach, and then
constraining the model parameters with a $\chi ^2$ minimization over a three
parameter space: effective temperature T$_{eff}$, surface gravity $log g$, and
metallicity [Fe/H].  Increasing the number of photometric measurements and
widening the wavelength coverage makes the fit more robust and reliable.

The T$_{eff}$ is the most stringently constrained parameter, with
uncertainties of order of $\sim$ 200\,K.  The other parameters $log g$ and
[Fe/H], were not well constrained and usually presented flat probabilities
distribution in the Bayes analysis.  Nevertheless, all the sources were fitted
with $log g\geq$4, and [Fe/H] $\geq-$1, the only exception was 2MASS
J15463089-5258367, which was better fitted with lower metallicity values
(-2$\leq$[Fe/H]$\leq-$1.5).  The photometry used in the SED fit is available
in the online version of Table \ref{Tab:phot}.

In order to use the optical catalogs GSC2.3, USNO-B1,UCAC4 and CMC15
\citep{Lasker2008,Monet2003,Zacharias2013,Muinos2014}, we had to make the following
assumptions: The first and second Blue/Red photographic band from GSC2.3 were
used as Johnson B/R bands (when two epochs were available for a given band we use the mean).
For USNO-B1 we used B$_j$ as Johnson B band and V mag as Johnson V.  For UCAC4
catalog, when the source had only R band measurement, we used the UCAC R
filter from VOSA, but, in the last release of UCAC4, they do not list the R
magnitude, they transformed the old optical magnitudes to Johnson B and V and
to Sloan r, i filters, and we adopted those values and filter systems.  For
these three catalogs we assigned a photometric error of 0.3 mag.  CMC15 uses
the SDSS r filter, the difference with the Sloan r system is of the order of
$0.02$mag, and a typical error of 0.08 mag at magnitude r$\sim$16 is listed in
the documentation
file\footnote{\url{http://svo2.cab.inta-csic.es/vocats/cmc15/docs/CMC15_Documentation.pdf}}.
To be safe we assumed a fixed value of 0.1 mag for the error in the
photometry.

We performed several simulations letting the uncertainties of USNO-B1, GSC2.3
and UCAC4, vary between 0.2 and 0.5 mag, and removing some of the photometric
points to see how much do these assumptions affect the final fit.  The SED
parameter change within the uncertainties listed above
$\Delta$T$_{eff}\sim$100\,K, $\Delta logg\sim$0.5, $\Delta logg\sim$0.5.

\begin{table}
\caption{Archival photometry for sources selected for EFOSC2
  spectroscopy and candidate companions}  
          
\label{Tab:phot}      
\centering   
\begin{tabular}{@{               }  l@{          }  l@{}   }        
\hline\hline                
 Column name\,\,\,\,   &  Description.  \\        
\hline                        
2MASS              & Name from 2MASS point source catalog \\ 
Spectra            & Spectroscopy available for this object   \\
B                  & B  from GSC2.3 or USNO-B1 \\ 
V                  & V  from GSC2.3 or USNO-B1 \\ 
R                  & R  from GSC2.3 or USNO-B1 \\
I                  & I  from Super Cosmos I$_{IVN}$ \\
R$_{UCAC}$         & old R  from UCAC4 \\
J$_{2MASS}$        & 2MASS J     \\
e\_J$_{2MASS}$     & Error in J$_{2MASS}$ \\
H$_{2MASS}$        & 2MASS H     \\  
e\_H$_{2MASS}$     & Error in H$_{2MASS}$ \\
K$_{S_{2MASS}}$    & 2MASS K$_S$ \\  
e\_K$_{S_{2MASS}}$ & Error in K$_{S_{2MASS}}$ \\
W1                 & WISE W1     \\ 
e\_W1              & Error in W1 \\
W2                 & WISE W2     \\     
e\_W2              & Error in W2 \\
W3                 & WISE W3     \\
e\_W4              & Error in W3 \\
W4                 & WISE W4     \\
e\_W4              & Error in W4 \\
I$_{DENIS}$        & DENIS I     \\
e\_I$_{DENIS}$     & Error in I$_{DENIS}$ \\
J$_{DENIS}$        & DENIS I     \\        
e\_J$_{DENIS}$     & Error in J$_{DENIS}$ \\
K$_{S_{DENIS}}$    & DENIS I     \\
e\_K$_{S_{DENIS}}$ & Error in K$_{S_{DENIS}}$ \\
$u_{SDSS}$         & SDSS u      \\        
e\_$u_{SDSS}$      & Error in $u_{SDSS}$ \\
$g_{SDSS}$         & SDSS g      \\           
e\_$g_{SDSS}$      & Error in $g_{SDSS}$ \\
$r_{SDSS}$         & SDSS r  or new UCAC4 r $^*$ or CMC15 r $^{**}$\\
e\_$r_{SDSS}$      & Error in $r_{SDSS}$ \\
$i_{SDSS}$         & SDSS i or new UCAC4 i $^*$   \\        
e\_$i_{SDSS}$      & Error in $i_{SDSS}$ \\
$z_{SDSS}$         & SDSS z    \\             
e\_$z_{SDSS}$      & Error in $z_{SDSS}$ \\
Y$_{UKIDSS}$       & UKIDSS Y  \\     
e\_Y$_{UKIDSS}$    & Error in Y$_{UKIDSS}$ \\
J$_{UKIDSS}$       & UKIDSS J  \\ 
e\_J$_{UKIDSS}$    & Error in J$_{UKIDSS}$ \\
H$_{UKIDSS}$       & UKIDSS H  \\
e\_H$_{UKIDSS}$    & Error in H$_{UKIDSS}$ \\
K$_{UKIDSS}$       & UKIDSS K  \\
e\_K$_{UKIDSS}$    & Error in K$_{UKIDSS}$ \\
Z$_{VVV}$          & VVV Z     \\       
e\_Z$_{VVV}$       & Error in Z$_{VVV}$ \\
Y$_{VVV}$          & VVV Y     \\        
e\_Y$_{VVV}$       & Error in Y$_{VVV}$ \\
J$_{VVV}$          & VVV J     \\       
e\_J$_{VVV}$       & Error in J$_{VVV}$ \\
H$_{VVV}$          & VVV H     \\      
e\_H$_{VVV}$       & Error in H$_{VVV}$ \\
K$_{S_{VVV}}$      & VVV K$_S$ \\
e\_K$_{S_{VVV}}$   & Error in K$_{S_{VVV}}$ \\
\hline                                   
\end{tabular}

 SDSS (DR9), GSC2.3, USNO-B1, UCAC4, CMC15, DENIS, VVV (DR3), 
2MASS, WISE \citep{SDSS2012,Lasker2008,Monet2003,Zacharias2013,Epchtein1997,Hempel2014,Skrutskie2006,Wright2010}.

* Magnitudes from UCAC4 transformed to SDSS r and i filters, we assume a fixed error of 0.3 mag.

** Magnitudes from CMC15, the catalog mentions they use a SDSS r filter, we assumed 0.1 errors (see discussion of errors in the text).
This table is available in a machine-readable form in the online
journal. Headers and description of the columns are shown here for guidance
only.

\end{table}

In Table \ref{Tab:dists} we provide the object 2MASS name, spectral type from
our EFOSC2 spectra and the uncertainty, and the effective temperatures
obtained through SED fit using VOSA and archival photometry, using the
BT-Settl models (log g and metallicities were free parameters and were almost
always above 4.5 and -1, respectively)\footnote{The only two objects that were
  automatically fitted by lower metallicities, 2MASS J1457-3904 and
  2MASS\,J15463089-5258367, are a known sub-dwarf and a early K star.}

To compare the spectral and the SED results we converted the spectral types
into effective temperatures using the relations of
\cite{Pecaut2013}\footnote{We used the table version 2014.09.29, available at
  \url{http://www.pas.rochester.edu/~emamajek/EEM_dwarf_UBVIJHK_colors_Teff.txt}}. 
We found a typical agreement within $\sim$100\,K between the two methods.

\begin{table*}
\caption{Spectral classification and spectro-photometric distances for the stars observed with EFOSC2@NTT.}     
\label{Tab:dists}     
\centering                         
\begin{tabular}{c@{                       } c@{      }  c@{             } c@{        }  c@{        } c@{    }  c@{         } c@{         }  c@{     }  c@{             }}        
\hline\hline                 
 2MASS Name   &  Sp. type & SED T$_{eff}$ &  J  & $\mu_{\alpha}$cos($\delta$)[mas] & $\mu_{\delta}$[mas] & d[pc] (lit.) & d[pc](this work) & ref.\\ 
\hline                       
J06571510-1446173 &  M4   $\pm 0.5$ & 3300  &   10.678   &    70   &    -270   &    --        & 29.1      &    this work$^{a}$ \\  
J07523088-4709470 &  M4.5 $\pm 0.5$ & 3200  &   11.738   &  -109   &     176   &    44.1$^{b}$& 38.2      &    4 \\
J08291581-5850305 &  K7   $\pm 1$   & 4200  &   10.206   &   382   &    -74    &    --        & 80.3      &    5 \\
J09432908-0237184 &  M4   $\pm 0.5$ & 3200  &   10.869   &  -200   &    -95    &    --        & 31.0      &    4 \\
J10570299-5103351 &  M2   $\pm 0.5$ & 3500  &   11.155   &  -617   &     78    &    54.1      & 73.0      &    6 \\ 
J11161471-4403252 &  M2.5 $\pm 0.5$ & 3600  &   10.649   &  -492   &    -3     &    --        & 52.2      &    1 \\
J11163668-4407495 &  M3.5 $\pm 0.5$ & 3300  &    9.917   &  -518   &    -29    &    21.5$^{b}$& 25.6      &    2 \\ 
J12412819-6507578 &  M4   $\pm 0.5$ & 3100  &   10.526   &  -499   &    -081   &    --        & 26.2      &    1 \\ 
J13211484-3629180 &  M5   $\pm 0.5$ & 2900  &   12.136   &  -513   &    -209   &    38.4      & 37.7      &    6 \\
J13322604-6621419 &  M4   $\pm 0.5$ & 3200  &   10.826   &  -294   &     226   &    --        & 31.4      &    1 \\
J13552455-1843080 &  M4   $\pm 0.5$ & 3200  &   14.042   &  -317   &    -96    &    --        & 136.1     &    3 \\
J14033647+0412395 &  M4.5 $\pm 1$   & 3000  &   15.853   &  -233   &    -073   &    --        & 250.3     &    7 \\
J14035016-5923426 &  M3   $\pm 0.5$ & 3500  &   10.258   &   11.5$\pm$5.1 & -492.2$\pm$4.3  & -- &  20.4$^{+3.4}_{-2.6}$ &  this work \\ 
J14040025-5923551 &  M2.5 $\pm 0.5$ & 3600  &   10.219   &    8.3$\pm$5.9 & -494.5$\pm$5.1  & -- &  24.8$^{+5.4}_{-3.7}$ & this work  \\
J14233830+0138520 &  M4   $\pm 1$   & 3000  &   12.374   &  -221   &    -195   &    --        & 52.8      &    7 \\
J14574906-3904511 &  M0   $\pm 1$   & 3900  &   13.693   &  -121   &    -405   &   215.6      & 328.6     &    6 \\
J15463089-5258367 &  K1   $\pm 1$   & 4700  &    8.737   &  -217   &    -199   &    44$^{b}$;77$^{b}$;56$^{b}$ & 71.3 & 8,9,10 \\
J15464497-5254371 & --              & 3100  &   12.340   &  -296.6$\pm$0.8 & -109.8$\pm$0.9 & -- &  42.6$^{+3.6}_{-3.0}$ &  this work \\
J15480441-5810533 &  M3.5 $\pm 0.5$ & 3400  &   10.169   &  -503   &    -207   &   33.9$^{c}$ & 29.5    &   1,10 \\ 
J17345391-6206546 &  K3   $\pm 1$   & 4800  &   10.364   &  -203   &    -390   &    --        & 131.5     &    1 \\
J19104599-4133407 &  M4   $\pm 0.5$ & 3300  &   10.610   &    68   &    -735   &   22.8$^{d}$& 27.6      &    2 \\
J19242108-0804516 &  M2   $\pm 0.5$ & 3700  &   10.766   &  -196   &    -379   &    --        & 59.8      &    1 \\
J20044356-7123334 &  K4   $\pm 1$   & 4500  &   10.168   &    80   &    -360   &    --        & 108.2     &    3 \\
J21252081-3422144 &  M4   $\pm 0.5$ & 3300  &   10.895   &  - 35   &    -450   &    --        & 32.3      &    1 \\
J22275385-2337300 &  M2   $\pm 1$   & 3600  &   14.422   &  - 62   &    -176   &    --        & 322.0     &    3 \\
\hline                                  
\end{tabular}

The Photometric distance error from this work are $\lesssim$20$\%$. 
$^{a}$: we recalculated the proper motion using 2MASS and the WISE ALL-SKY epoch position
with the highest S/N, as we find that the \cite{Salim2003} value for proper
motion of LP 721-15 is inconsistent with the motion of the sources in the
images, this object could be an unresolved binary, and then located furher
away, see text for discussion; $^b$: photometric distance (spectral type derived from
T$_{eff}$ using the relations in \cite{Pecaut2013} and on-line table maintained
by E. Mamajek, see text for the link); $^{c}$: parallax distance; $^{d}$: the paper
cites the value for other object of the system (SIPS 1910-4133A) and is from
photographic plates, our measurement is for SIPS 1910-4133B.  The values of proper motion and distances with
quoted errors were fitted from VVV data: (1)~\cite{Luhman2014a} ; (2) \cite{Winters2015}; (3)
\cite{Salim2003} ; (4) \cite{Finch2007}; (5) \cite{Luyten1980}; (6)
\cite{Subasavage2005a}; (7) \cite{Lepine2005a}; (8) \cite{Ammons2006} (9)
\cite{Fresneau2007};(10) \cite{Pickles2010}
\end{table*}

\section{Distance estimation}

\label{Sec:dist}
\subsection{Spectro-photometric distances}
\begin{figure}
   \centering
   \includegraphics[width=\columnwidth]{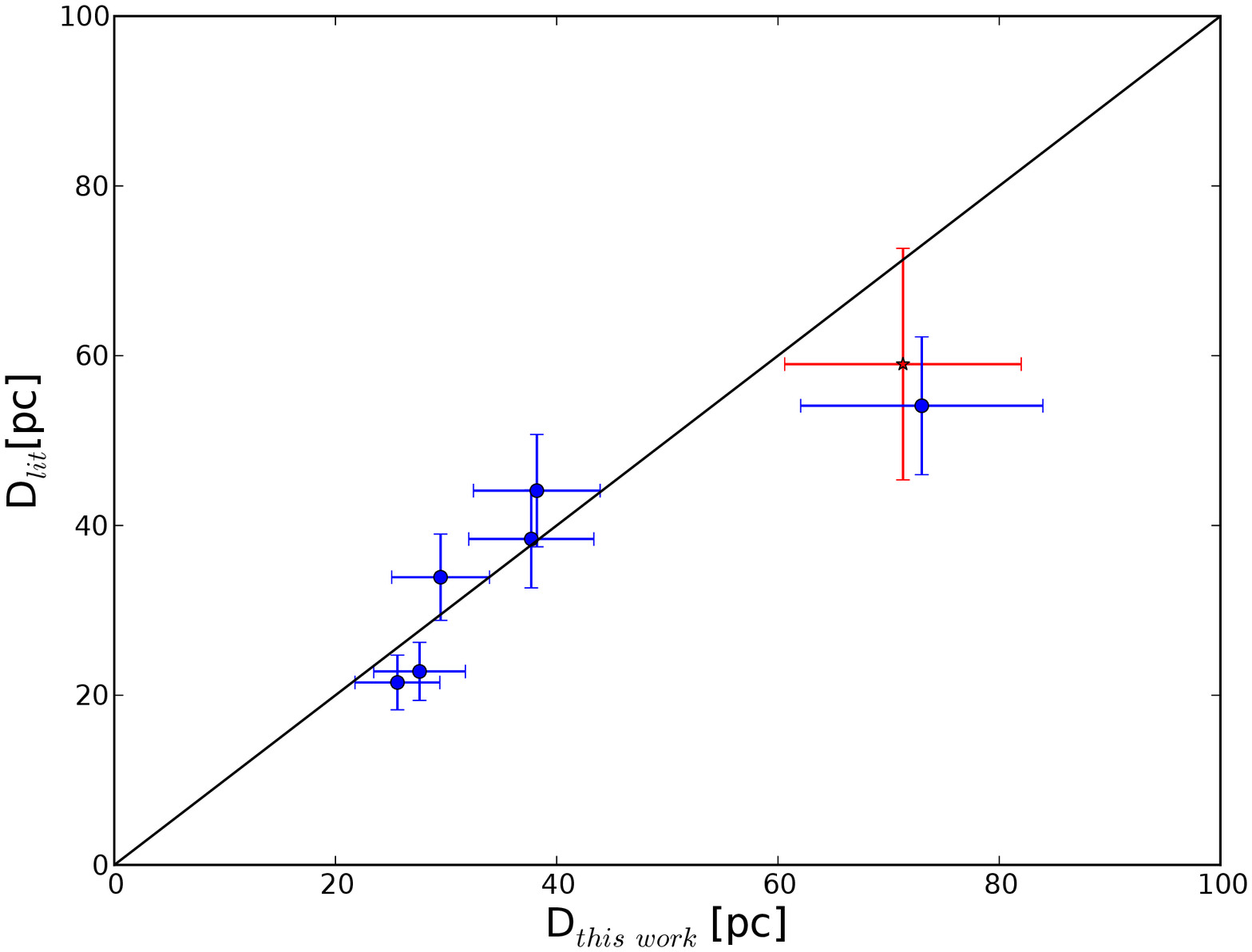}
      \caption{Literature photometric distances v/s spectro-photometric distances 
      from this work for objects within 100\,pc. 
      The red star represents the average of the three distance from the 
      literature for source 2MASS J1546-5258, and the error in the Y axis for this source 
      is the dispersion around the mean.
      The solid black line shows a 1:1 relation, and not a fit to the data points.}
\label{fig:compare_dist}
\end{figure}

We estimated the distances to the objects and their companions from the
derived spectral types and T$_{eff}$ from the 2MASS $J$ and $K_S$ photometry
using the absolute magnitudes from \cite{Pecaut2013}.  For each object we
calculated the distances for the two bands separately.  The mean difference is
3.1$\% \pm$1.7$\%$, with a maximum value of 5.6$\%$ for object
2MASS\,J1403+0412.  Our final estimate was the average of the two
measurements.  The sub-type error classification affects the distance
estimation by a factor of 15$\%$, on average.  The photometric uncertainties
introduce errors within 1-3$\%$ in the magnitude range we examine.
Therefore, estimated spectro-photometric distances will have an associated
error of $\lesssim$20$\%$, of the same order as the differences between our
measurements and the values available in the literature (Table
\ref{Tab:dists}).
 A comparison between the values for the distances in the literature and
 the values we obtain in this study is shown in Fig. \ref{fig:compare_dist}. The 
 error bars correspond to 15$\%$ of their distances in each axis. The only exception
 is  2MASS J1546-5258, as this source has three distances estimates in the literature,
 We plotted the average value of the distance and quote the standard deviation around 
 the mean as the error.

\subsection{Parallax distances}
Four objects in our list lie within the VVV survey footprint: 2MASS
J14035016-5923426, 2MASS J14040025-5923551, 2MASS J15464497-5254371, 2MASS
J15463089-5258367, but the last one is badly saturated in $K_S$, and reliable
positions could not be determined.  Therefore, we measure the parallaxes only
for the first three sources using the procedure developed in the Torino
Observatory Parallax Program described in \cite{Smart2003}. This has been
adapted to the format of the VVV data products (coordinates and photometry)
delivered by CASU 
\footnote{\url{http://casu.ast.cam.ac.uk/surveys-projects/vista/data-processing}}.
A detailed description of the parallax code can be found in that paper, and we
give here only a brief description of the steps involved. For the purpose of
the astrometric reduction, we selected reference stars in a circle of radius
1\arcmin\ around the target which, given the high stellar density in these
fields, provides an adequate number of reference stars (above a hundred
sources). These were selected among the highest S/N objects in the field of
view, satisfying the condition that they appear in at least 80\% of the
frames, and do not exhibit large proper-motions.

\begin{figure*}
   \centering
   \includegraphics[width=\columnwidth]{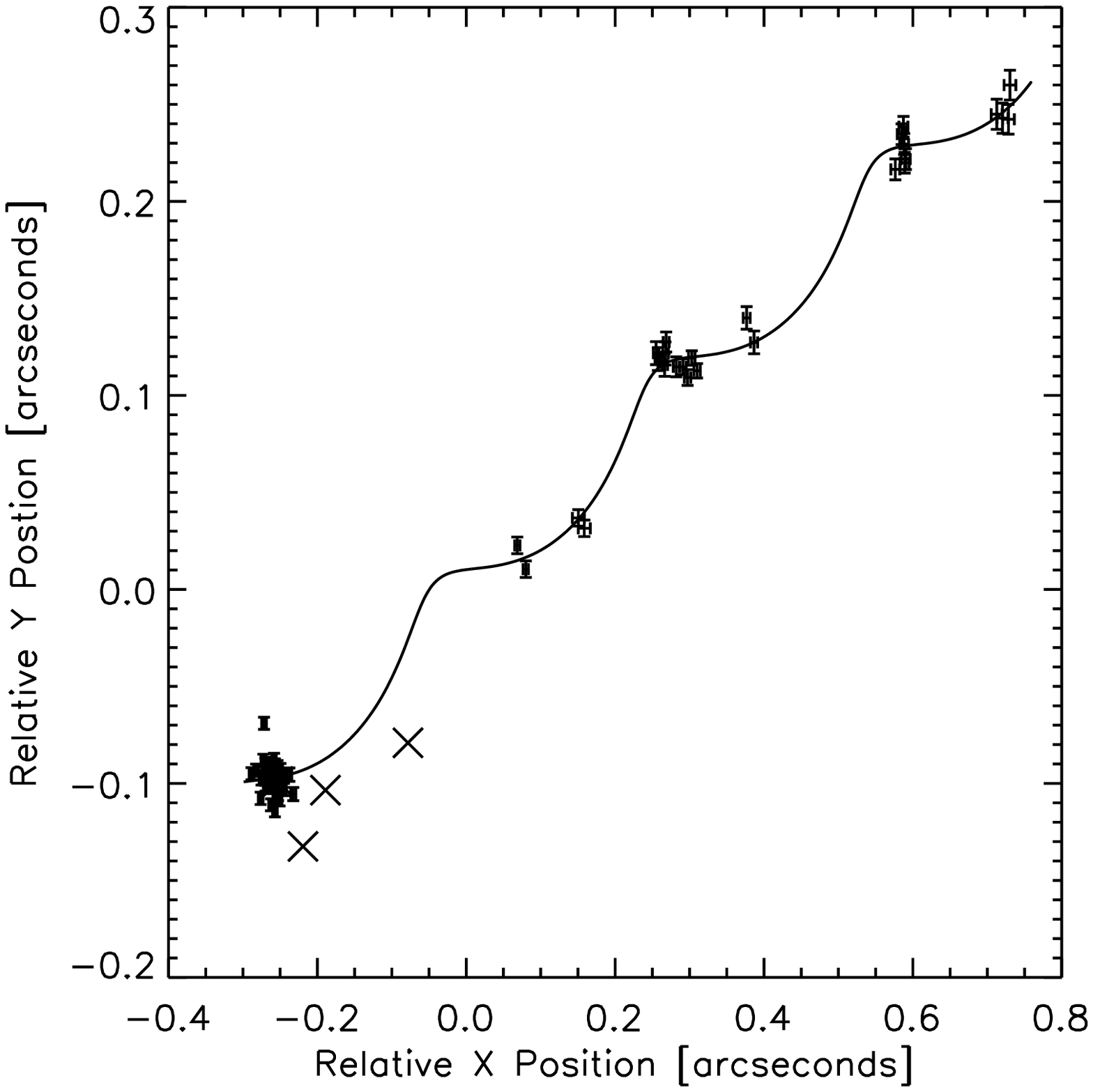}
   \includegraphics[width=\columnwidth]{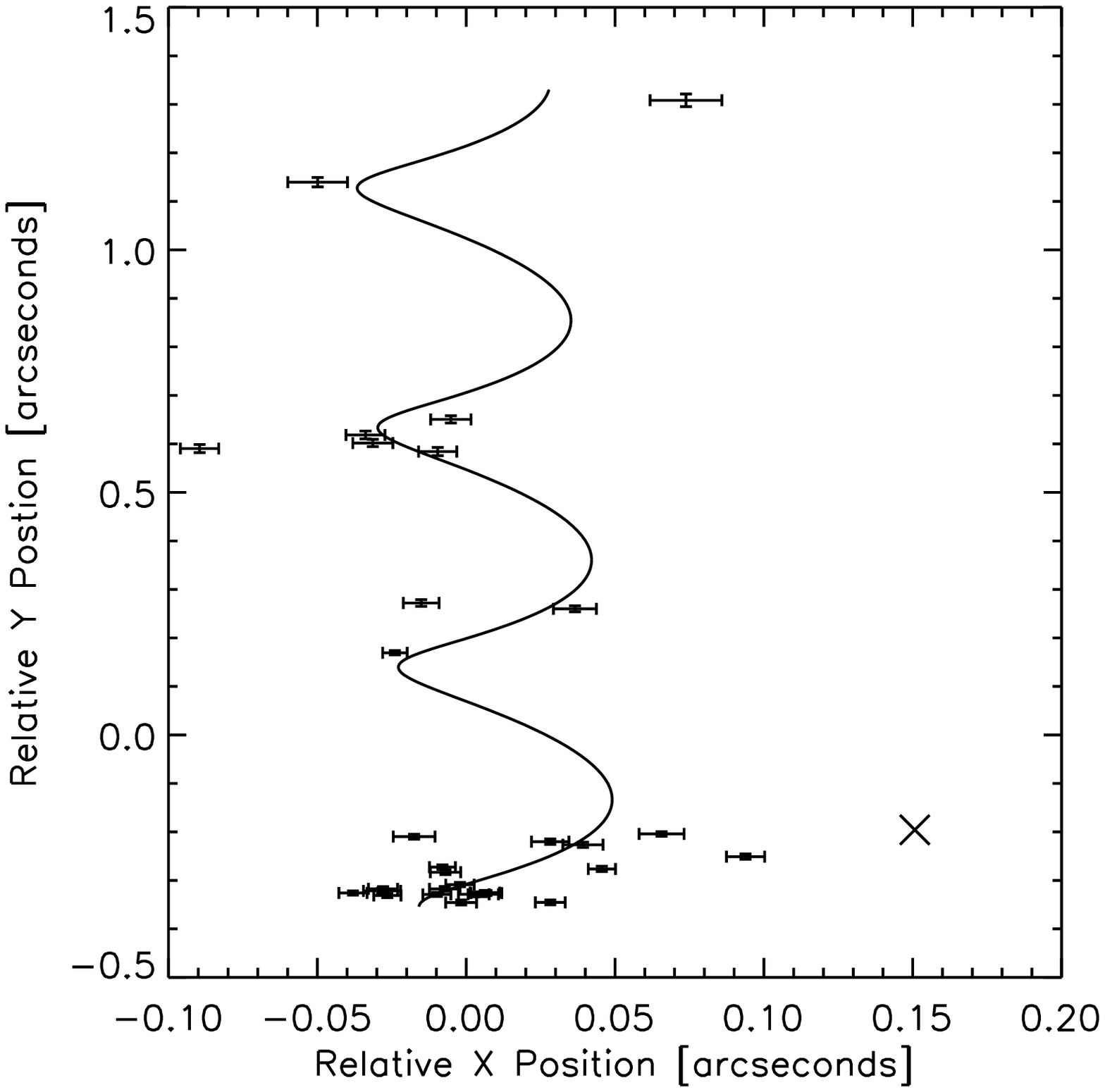}
   \includegraphics[width=\columnwidth]{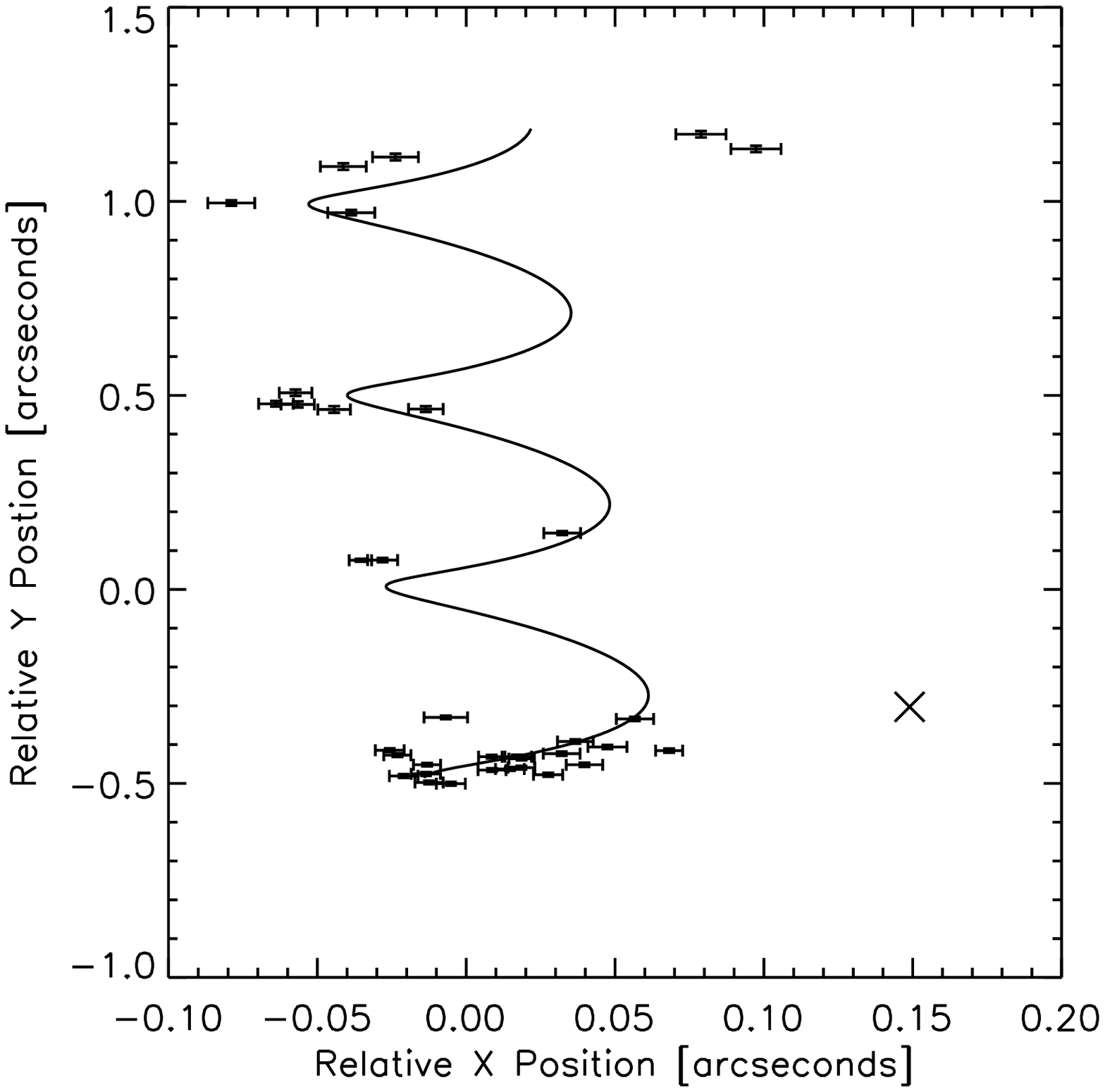}
      \caption{  Relative displacement of the targets 2MASS 15464497-5254371, 2MASS
        J14040025-5923551 and 2MASS J14035016-5923426 with respect to the
        adopted reference frame as a function of time (top of each plot epoch
        2010, bottom of each plot, epoch 2013).  A base-frame was chosen for
        the registration of all other frames to account for small offsets and
        rotation between them.  The fitted parallax wobble, shown by the solid
        line, is clearly seen in the observations which are plotted as crossed
        error bars representing the formal VISTA pipeline errors. Any rejected
        observations within the plot ranges are plotted as Xs.}
\label{fig:par}
\end{figure*}

For example, for 2MASS 15464497-5254371 the initial number of reference stars
was 189, but in the end only 121  of these were used to build the astrometric
reference frame. In total we had 70 VVV images in $K_s$, at various epochs
(see Figure \ref{fig:par}), spanning more than 3 years.  Four epochs were
excluded from the solution due to their high residuals with respect to the
mean solution. Despite the relatively small parallax, the long baseline and
the parallax factor coverage provide a small final  error of 1.80\,mas. The
conversion from relative to absolute parallax, which in any case is quite
small (less than 0.4\,mas), was computed using the Galactic model by
\cite{Mendez1996}, extended to IR-wavelengths.  The proper motion and
parallaxes for these three sources are listed in Table \ref{Tab:dists}, and
Fig. \ref{fig:par} show the  observations of each source and the best fits.

 The binary system 2MASS J14035016-5923426 and 2MASS J14040025-5923551 was on
the border of two adjacent chips in the observational sequence and  both objects 
are very bright often saturating in good seeing.  This led to
high centroiding errors and a sparse reference  star set of only 41 and 55 objects
for the two fields respectively. The two parallax solutions were therefore
completely independent with different reference fields and with observations
from different VIRCAM  chips.

Each observation has a quote positional error from the VISTA pipeline  but
 there is a significant fraction of the total that is systematic 
in nature from transforming the observations to a common system. For this reason 
when calculating the errors of the derived parallax we do not use the individual 
observation errors.  The final quote errors on the target parameters are obtained from 
the covariance matrix of the solution scaled by the error of unit weight.
 
The observations used in the sequence are selected using the standard outlier
rejection criteria developed in the Torino program following these two criteria:
1) the average per coordinate error of the reference stars in a frame  must be less than 
the mean error of all frames plus three standard deviations about the mean;
2) The combined observed-minus-computed coordinate residual of each observation
 must be less than three times the sigma of the
whole solution. The objects 2MASS J15464497-5254371, 2MASS J14035016-5923426
and 2MASS J14040025-5923551 had 4, 2 and 2 observations rejected respectively
by these criteria.

Extensive bootstrap-like testing was carried out on the observations to make
sure the results were robust. This consisted of  iterating through each
observation and using as the primary base frame and thus making a solution that that
incorporated slightly different sets of reference stars and a different
starting point within the sequence. The solution chosen for publication is that one 
which is closest to the median of all solutions. The majority
of the solutions  ($>$90$\%$) were all within one sigma of the chosen solution. 

\section{Discussion and remarks on individual objects}
\label{sec:discussion}

We determined physical properties and distances for twenty five stars, most of
which are within D$\sim$140\,pc from the Sun. Only eight of these had
previously determined distances.  Thirteen of the stars are located within
D$\sim$50\,pc. Given that our selection of targets from the WISE high PM list
was not systematic, and that the list itself is not a complete census of the
objects within D$\sim$50\,pc we did not attempt to do a completeness analysis.
The derived spectral types range from early-K to mid-M.

The high transverse velocities of 2MASS J173453.91-620654.6, 2MASS
J140336.47+041239.5 and 2MASS J145749.06-390451.1 and the co-moving binary
pair 2MASS J08291581-5850305 (or L 186-122) and 2MASS J08292286-5849209, make
them likely members of the Galactic halo.  The remaining objects probably
belong to the disc population.

\subsection{Probable multiple systems}
We adopted binarity criteria requiring: common PMs i.e. $\Delta \theta \Delta
\mu \leq (\mu/0.15)^{3.8}$\,, where $\Delta \theta$ and $\Delta \mu$ are the
angular separation (in arcseconds) and the difference in the magnitude of the 
proper motion vectors (in arcseconds per year) \citep[See Sect. 2.2. in ][]{Lepine2007}, 
distances in agreement at the 2$\sigma$ confidence level, and consistency between 
the spectral types and the apparent magnitudes.

Based on these criteria we identify six probable multiple system, five
previously reported systems, and discovering one new one.  The following stars
are most likely real gravitationally bound systems:

\begin{itemize}
 \item 2MASS J06571510-1446173 (or LP  721-15) and 2MASS J06571773-1446382; 
 \item 2MASS J08291581-5850305 (or L 186-122) and 2MASS J08292286-5849209;  
 \item 2MASS J14040025-5923551 (or L 197-165) and 2MASSJ14035016-5923426; 
 \item 2MASS J15480325-5811119 (or LHS 3119) and 2MASS J15480441-5810533; 
 \item 2MASS 19103460-4133443  (or SIPS1910-4133A), 2MASS 19104599-4133407(or SIPS1910-4133B) and 2MASS 19103359-4132505 (or  SIPS1910-4132C);
 \item 2MASS J20044356-7123334 (or LTT 7914) and 2MASS J20043661-7123532.
\end{itemize}

Their parameters are listed in Table \ref{Tab:bins}.

The triple system 2MASS 19103460-4133443, 2MASS 19104599-4133407 B and 2MASS
19103359-4132505 C was reported by \citep{Lepine2005b,Deacon2005,Hambly2005}.
The last work discussed the possibility that this is actually a quadruple
system because the magnitude of the component B is $\sim$0.5 mag brighter than
the C component. The spectro-photometric and SED based distances all agree
within the 1$\sigma$ errors (see Table \ref{Tab:bins}. When we compare the
T$_{eff}$ obtained via SED fit of components B and C we find a difference of
$\sim$200\,K which would be sufficient to explain the 0.5 mag difference (that
difference can be less given the error bars for temperature estimates). But we
think that is not necessary to invoke a possible equal mass unresolved binary
in component B to explain that difference in magnitude, and it is more likely
to be explained as an effect of slight difference ($\sim$100-200\,K) in
T$_{eff}$.

2MASS J14040025-5923551 and 2MASSJ14035016-5923426 are co-moving.  Our spectra
suggest this is a M3+M2.5 nearly equal mass binary. Despite the saturation, we
were able to measure a parallax from the multiepoch VVV data:
40.35$\pm$7.19\,mas and 49.07$\pm$7.06\,mas (24.78$^{+5.4}_{-3.7}$\,pc
and 20.4$^{+3.4}_{-2.6}$\,pc) respectively.

2MASS J20044356-7123334 (or LTT 7914) was observed by the RAdial Velocity
Experiment (RAVE) in its fourth data release \cite{Kordopatis2013} and
described there.  They derived a T$_{eff}$=4817 and $log$\,$g$=4 and
metallicity [Fe/H]=0.05, based on this we could assume the dwarf nature and
apply the relations described before, deriving a distance of 108\,pc.  They
also obtained a radial velocity of
RV$_{Heliocentric}$=$-$50.2$\pm$2.2\,km\,s$^{-1}$ Our best spectral type for
this object is K2, the reference T$_{eff}$ for a K2V star is $\sim$5000\,K so our
classification might be revised by one sub-type.  Applying photometric SED
fitting we obtained a T$_{eff}$=4500\,K$\pm$100, $log$\,$g$=5$\pm$1 and
metallicity [Fe/H]=0.3$\pm$0.5 The photometric spectral type would be K4,
which is 2 spectral types later than the fit from our optical spectroscopy
and one later than the type inferred by  RAVE. For a range of photometric
distances for K2-K4 types we obtain distances of 100-130 pc, implying a
tangential velocity of 170-225 km\,s$^{-1}$, typical for thick disk or halo
objects. Interestingly, the available data does not support a low
metallicity for this object.  For the co-moving star 2MASS J20043661-7123532
the best photometric fit was T$_{eff}$=3100\,K$\pm$100, $log$\,$g$=5.5$\pm$1
and metallicity [Fe/H]=-1$\pm$0.5.  Assuming a spectral type of M4-M5, the
distance would be 80-120\,pc in agreement within the errors to the estimated
value for 2MASS J20044356-7123334 The objects are separated by 38.7\arcsec on
the sky, which corresponds to $\sim$4200\,AU for a distance of 110 pc.

Object 2MASS J22275385-2337300 (or LP 876-22) was observed by mistake, as the
real new binary candidate was LP 876-1 and 2MASS J22274199-2337283.  The
observed target was classified as M2$\pm$1 star, if we compute the distance we
obtain 322\,pc which will put this object in a tangential velocity over 250
km/s and hence probably this object belongs to the halo population, but the
distance might be considerably less if we consider that this object might be
metal poor, as happened to be with previous sources.  
We perform the SED fitting to the co-moving pair LP 876-1, 2MASS J22274199-2337283 
and they were classified as a M3.5-M8.5, that would imply a distance between
40\,pc-50\,pc, but further observations are required to settle their true
nature.

\subsection{Rejected multiple systems}
The following objects are not real physical pairs, the argument for rejecting
them as binaries are: the total proper motion, position angle of motion,
spectral types compared to photometric spectral type and distances estimates
for the primary and secondary, do not agree within the expected errors.
\begin{itemize}
\item 2MASS J07523088-4709470 and 2MASS J07523777-4717270;
 \item 2MASS J09432908-0237184 and 2MASS J09434389-0229570;
 \item 2MASS J10570299-5103351 and 2MASS J10573037-5102190;
 \item 2MASS J11163668-4407495 (or LHS 2386) and 2MASS J11161471-4403252 (see text);
 \item 2MASS J13211484-3629180 and 2MASS J13214404-3627316;
 \item 2MASS J13552455-1843080 (or LP 799-1) and 2MASS J13553933-1840586;
 \item 2MASS J14033647+0412395 and 2MASS J14040651+0418532;
 \item 2MASS J14233830+0138520 and 2MASS J14234208+0146235;
 \item 2MASS J14574906-3904511 and 2MASS J14582414-3907504;
 \item 2MASS J15463089-5258367 and 2MASS J15464497-5254371.
\end{itemize}

2MASS J07523088-4709470 and 2MASS J07523777-4717270: The second object moves
$\sim$2 times faster than the first object ($>$5$\sigma$ outlier) and the
derived T$_{eff}$ of the secondary (the fainter source) is 200\,K higher.
Given the classification of M4.5V for the primary, this would place the
secondary at least twice as far.

2MASS J11163668-4407495 (or LHS 2386) and 2MASS J11161471-4403252 these were
classified and found to be a co-moving pair observed by G.P. Kuiper and
re-classified in \cite{Bidelman1985}, he classified LHS 2386 as M3: We obtain
a best fit with M3.5V and M2.5 for 2MASS J11161471-4403252.
\cite{LuhmanShep2014} list these sources as a co-moving binary, because the
difference in total proper motion and position angle is small (4.5$\%$ and
1.6$\%$ respectively). Also the probability of being a chance alignment is
below 10$\%$ based on the criteria from \cite{Lepine2007}. However, the
distance we derive for the two stars are only consistent at 3$\sigma$ level. We
obtain half the distance for LHS 2386 than for 2MASS J11161471-4403252. Even
if 2MASS J11161471-4403252 is an equal mass binary that would place it around
44-50\,pc $\sim$10-15\,pc farther away than we expect for LHS 2386. In this
hypothetical case, the distances will agree within the errors, that would mean
that at 5.9\arcmin of angular separation and distances between 44 and 50\,pc,
the projected physical separation would be $\sim$15-20 thousand AU which would
place them as one of the scarce population of very low mass and very wide
binaries within 50\,pc. Radial velocities of both stars and a more robust
estimation of their distances are needed to disentangle their true nature, but
we do not list it as a physical pair with the present evidence.

2MASS J14574906-3904511 and 2MASS J14582414-3907504 have PMs that agree within
1$\%$ when comparing the 2MASS and WISE positions, and the position angle of
the motion differ only by one degree, these stars should probably be a real
physical pair. On the other hand the inferred spectro-photometric distance for
the primary is 328.7\,pc, while for the secondary using the SED fit we obtain
 a distance around 40\,pc for spectral type M7. In \cite{Jao2008} they 
found that the primary object is actually a MI.0VI sub-dwarf and the 
distance derived by \cite{Subasavage2005a} is 215.6\,pc even assuming this shorter distance the
object is not consistent with a real binary and would imply a high tangential 
velocity of 432 km/s.  Finally, the angular separation of
445.9\arcsec means that the projected physical separation at 215.6\,pc would
be 0.466\,pc.  Combining these two arguments we argue this is not a real
binary. 

2MASS J15463089-5258367 and 2MASS J15464497-5254371 \cite{Ammons2006} derived
a distance 44$^{+30}_{-15}$\,pc and estimated a T$_{eff}$=4669-4754\,K
(according to different fitting functions) based on Hipparcos (Tycho) data for
the first object. Other two attempts to measure the distance from photometry
are available from \cite{Fresneau2007} and \cite{Pickles2010} they obtained
56\,pc (no error bars) and 77$^{+50}_{-22}$\,pc respectively. Our best SED fit
yields 4700\,K, in agreement with \cite{Ammons2006}, but our best spectrum fit
is between K0V-K2V. If we assume this as the correct spectral type, then the
photometric distance we obtain is between 68-77\,pc, for K0-K2 respectively.
For the second object we were able to derive a distance based on parallax from
VVV, as discussed in the previous  section $\pi$= 23.5$\pm$1.8 mas
(42.6$^{+3.6}_{-3.0}$\,pc). The parallax and proper motion of 2MASS
J15464497-5254371 are shown in Figure \ref{fig:par}  and Table \ref{Tab:dists}. 
The distance agrees very well with the value derived by \cite{Ammons2006}, but 
is almost 3$\sigma$ away from the photometric distance to 2MASS J15463089-5258367,
in addition to the difference in the proper motion between the primary (from 
Tycho-2 catalog \citealt{Hog2000}) and our measurements for the candidate companion 
are too large. The evidence does not support that these two stars form a real binary,
the parallax and more accurate proper motions for both sources will be
measured by Gaia mission, and then the true nature of  these objects will be
settled.

\subsection{Possible future microlensing event}
While looking for the available photometry for the object NLTT 37178 from the
virtual observatory, we found a nearby source, classified as extragalactic
(photometric redshift 0.13) with photometry from SDSS and GALEX. In the
following years this object will be getting closer until the closest approach
to the center, with the closest approach of 0.6\arcsec\ in 4--8 years. Some
extended emission in the galaxy is visible on SDSS images, and we can expect
that the nearby star can act as a lens for the outskirts of this galaxy. Deep
U and B band pre-lensing observations are needed to characterize the background
source. Search for microlensing events during the next few years may be
promising. The most favorable filters to observe the galaxy will be U,B
(and/or UV filters from space).  The microlensing event might help to 
understand the real nature and physical properties of this object, e.g. 
if it is an unresolved binary or hosts a planet.  More robust distance
estimations (parallax) are necessary, to better constrain the Einstein radius
for the system. We assumed a mass of 0.2M$_\odot$ for the lens and distances
of 40\,pc and 543.9\,Mpc(co-moving radial distance\footnote{value obtained
  using the NED cosmological calculator
  \url{http://www.astro.ucla.edu/~wright/CosmoCalc.html}}) for lens and source
respectively, and obtain a crude estimate of the Einstein radius of
6.4\,mas. As the lensed source is resolved, we might expect variations on the
light curve due to lens magnifying different parts of the galaxy.

\section{Conclusions} 
\label{conclusions}
We performed spectroscopic follow up for over twenty new high proper motion
objects found by the WISE satellite, and looked for possible new wide binary
companions.  We found one T2 dwarf probably located within 15\,pc.  We
obtained optical spectral types and photometric distances for 24 objects, as
well as parallax measurements for 3 of them.  We present some additional
evidence for six co-moving objects that are likely physical pairs, two of them
are new binary candidates.  Four objects are probable members of the galactic
halo given their large tangential velocities.

Most of the objects analyzed in this study are located within 75\,pc from the
Sun, and are bright enough for further follow up and search for planets using
state of the art and upcoming NIR instruments.

The use of relatively loose constraints when selecting possible wide co-moving
companions given the in-homogeneity of resources available in  the literature
prove useful to find new co-moving stars. This causes many false
positives, but they can be eliminated a
posteriori using multiple arguments, combining proper motion,
physical separation and spectral energy distributions in the calculation of
photometric distances. It is also important to emphasize the relevance of
obtaining distances or spectral types for discriminating chance alignments
from real wide binaries (or co-moving stars),  even when the proper 
motion and position probabilities are very low we show two examples in this paper
where the hypothesized binaries are most likely not physically related.

We also discussed a likely microlensing event due to a star passing in front of
a background galaxy.  The number of predicted microlensing events of this type
will be more frequent as more HPM low mass objects are
found in high density environments like the galactic bulge and inner disk, but
also with background galaxies.  Although the lens candidate we present here
is not predicted to pass in front of the center of the galaxy the event 
can be used to study the lens for unresolved companions
and planets. In this case this maybe particularly difficult as the lens 
goes through different parts of the galaxy in short time scales magnifying 
regions of intrinsically different brightness.  These events can also be used to 
make more detailed structural studies of galaxies at low redshift.

\begin{table*}
\caption{Binary/Multiple system investigated in this paper. The columns are:
  spectral type are from this work and from literature spectra, T$_{eff}$ is
  defined based on photometric spectral energy distribution, 2MASS J band
  magnitude, $\rho$ the angular separation in arcsec, D is the photometric
  distance in pc, except for the labeled objects which have parallaxes,
  $\mu_{\alpha}$cos($\delta$) and $\mu_{\delta}$ are the proper motion in each
  celestial coordinate.}
\label{Tab:bins}      
\centering                          
\begin{tabular}{c@{     } c@{      } c@{        } c@{        } c@{       } c@{         } c@{          } c@{          } c@{                                  }}     
\hline\hline              
 Star ID   &  Sp. type  & T$_{eff}$ &  J$_{2MASS}$ & $\rho$& Dist.  & $\mu_{\alpha}$cos($\delta$) & $\mu_{\delta}$ & remarks \\  
  & (reference) & [K] & [mag] & [\arcsec] & [pc]   &  [mas]    &[mas]& \\
  \hline                  
\hline
\multicolumn{6}{l}{} \\
2MASS J06571510-1446173   & M4      & 3300  & 10.678 & 43.6  &  29.1       &   70      &    -270     &  see ** in table \ref{Tab:dists}\\                                  
\multicolumn{6}{l}{} \\
2MASS J06571773-1446382   & ---     & 3000 & 12.751 & ---   &  $\sim$ 40  &   61 (1)  &    -268 (1) &   \\                                  

\hline
\multicolumn{6}{l}{} \\
2MASS J08291581-5850305   & K7      & 4200  & 10.206 & 88.49 &  80.3      &  363 (4)  &    -81(4)   &  Halo wide binary\\                                  
\multicolumn{6}{l}{} \\
2MASS J08292286-5849209   & ---     & 2900 & 14.550 & ---   &  $\sim$110  &   355     &    -73(1)   &   \\                                  
\multicolumn{6}{l}{} \\
\hline
\multicolumn{6}{l}{} \\
2MASS J14040025-5923551 &  M2.5   & 3600   &  10.219 & 78.05   & 24.78$^{+5.4}_{-3.7}$   &  8.27$\pm$5.9 & -494.5$\pm$5.1   & VVV parallax \\                                  
\multicolumn{6}{l}{} \\
2MASS J14035016-5923426 &  M3     & 3500   & 10.258  & ---     & 20.4$^{+3.4}_{-2.6}$    &  11.5$\pm$4.6 & -492.2$\pm$4.3   &  \\                                  
\multicolumn{6}{l}{} \\
\hline
\multicolumn{2}{l}{} \\
2MASS J15480325-5811119   & M1.5(1) & 3467(7)  &  8.379 & 20.7  &  23.5(1)    & -593 (1)  &    -250 (1) & V$_{rad}$=30.9 km/s (1)\\                                  
                   &  &  &   &  & 33.6(2) &  &      & \\                                  
\multicolumn{6}{l}{} \\
2MASS J15480441-5810533   & M3.5    & 3400  & 10.169 & ---   &  29.5       & -503 (3)  &    -207 (3) &   \\                                  
\hline
\multicolumn{6}{l}{} \\
2MASS 19103460-4133443    & ---     & 3500  & 9.851 & 127.78(A-B) & 35.9   &  72 (5)  &    -738(5)   &  Triple system\\                                  
            &  &  &   & 54.99(A-C) &  &  &   & \\                                  
\multicolumn{6}{l}{} \\
2MASS 19104599-4133407    & M4      & 3300   & 10.610 & 147.90(B-C)& 22.8 (6)   &   91 (5)     &    -742 (5)   & Component B  \\                                  
            &  &  &   &  & 27.6    &  &      & \\
\multicolumn{6}{l}{} \\
2MASS 19103359-4132505    & ---     & 3100  & 11.147 & ---        & 29.2   &   68 (5)     &   -735 (5)   &  Component C \\  
\multicolumn{6}{l}{} \\
\hline

\end{tabular}

(1) \cite{Hawley1996}, (2) \cite{VanAltena1995}, (3) \cite{Luhman2014a} (4)
\cite{Kirkpatrick2014}, (5) \cite{Deacon2005}, (6) \cite{Winters2015}, (7)
\cite{Houdebine2010}.  The values without reference were calculated or derived
in this work, the photometric distances have a $\lesssim$20$\%$ error (spectral
type and effective temperature uncertainties are the main source of error).

\end{table*}

\section*{Acknowledgements}
The authors thank the referee Dr. Nigel Hambly for his comments and suggestions 
that helped to the improve the quality of the paper.
J.C.B., D.M., acknowledges support from: PhD Fellowship from CONICYT, Project
FONDECYT No. 1130196.  D.M and R.A.M acknowledges project support from Basal
Center for Astrophysics and Associated Technologies CATA PFB-06.  Support for
J.C.B, D.M, R.A.M, M.G and R.K is provided by the Ministry of Economy,
Development, and Tourism's Millennium Science Initiative through grant
IC120009, awarded to The Millennium Institute of Astrophysics, MAS.
A.B. Acknowledges financial support from the Proyecto Fondecyt de Iniciaci\'on
11140572.  Part of this work was completed at the ESO Headquarters, Garching
bei M\"unich, with support from the ESO Director Gerneral's Discretionary Fund
program.  R.A.M. acknowledges ESO/Chile for hosting him during his
sabbatical-leave throughout 2014.  MG acknowledges support from Joined
Committee ESO and Government of Chile 2014.  This research has made use of the
SIMBAD database, operated at CDS, Strasbourg, France This publication makes
use of data products from, the Two Micron All Sky Survey, which is a joint
project of the University of Massachusetts and the Infrared Processing and
Analysis Center/California Institute of Technology, funded by NASA and NSF.
This publication makes use of data products from the Wide-field Infrared
Survey Explorer, which is a joint project of the University of California, Los
Angeles, and the Jet Propulsion Laboratory/California Institute of Technology,
funded by the National Aeronautics and Space Administration.  Based on data
from CMC15 Data Access Service at CAB (INTA-CSIC).  This research has
benefitted from the M, L, T, and Y dwarf compendium housed at
DwarfArchives.org This research has benefitted from the SpeX Prism Spectral
Libraries, maintained by Adam Burgasser at
http://pono.ucsd.edu/~adam/browndwarfs/spexprism






\bsp	
\label{lastpage}
\end{document}